\documentclass[aps,pre,twocolumn,nofootinbib,floatfix,showpacs]{revtex4}  

\usepackage{color}
\usepackage{soul}

\usepackage{graphicx} 
\usepackage{amssymb}
\newcommand {\be}{\begin{equation}}
\newcommand {\ee}{\end{equation}}
\newcommand {\ba}{\begin{eqnarray}}
\newcommand {\ea}{\end{eqnarray}}

\newcommand{\To}{T_{_0}}
\newcommand{\DT}{\Delta T}
\newcommand{\Tl}{T_{_L}}
\newcommand{\Tr}{T_{_R}}
\newcommand{\Mmax}{M_{\mathrm{max}}}
\newcommand{\Mmin}{M_{\mathrm{min}}}

\begin{document}

\title{Thermal rectification in three-dimensional mass-graded anharmonic oscillator lattices}

\author{M.~Romero-Bastida}
\affiliation{SEPI ESIME-Culhuac\'an, Instituto Polit\'ecnico Nacional, Avenida Santa Ana No. 1000, Colonia San Francisco Culhuac\'an, Delegaci\'on Coyoacan, Distrito Federal 04440, Mexico}
\email{mromerob@ipn.mx}
\author{M.~Lindero-Hern\'andez}
\affiliation{Centro de Investigaci\'on en Ciencia Aplicada y Tecnolog\'\i a Avanzada-Quer\'etaro, Instituto Polit\'ecnico Nacional, Cerro Blanco 141, Colinas del Cimatario, 76090 Santiago de Quer\'etaro, Quer\'etaro, Mexico}
\email{miguel.lindero@gmail.com}

\date{\today}

\begin{abstract}
In this work we study the thermal rectification efficiency, i.e., asymmetric heat flow, of a three-dimensional mass-graded anharmonic lattice of length $N$ and width $W$ by means of nonequilibrium molecular dynamics simulations. The obtained rectification, which is of the same order of magnitude as that of the corresponding one-dimensional lattice, saturates at low values of the aspect ratio $W/N$, consistent with the already known behavior of the corresponding heat fluxes of the homogeneous system under analogous conditions. The maximum rectification is obtained in the temperature range wherein no rectification could be obtained in other one-dimensional systems, as well as in the corresponding one-dimensional instance of the model studied herein.
\end{abstract}


\maketitle

\section{Introduction\label{sec:Intro}}

Thermal rectification, i.e., asymmetrical heat flux, is a phenomenon that, although discovered a long time ago~\cite{Starr36}, only recently became an interesting research topic, mainly because of the results obtained by numerical simulations in one-dimensional (1D) systems of anharmonic oscillators~\cite{Roberts11,Li12,Wehmeyer17}. The first theoretical proposal of a thermal device which could rectify the heat current when the temperature gradient was reversed consisted of a three-segment nonlinear lattice with a Morse on-site potential with different parameters in each segment~\cite{Terraneo02}. However, its gain, defined as the ratio of thermal fluxes in two opposite directions, was found to be only a factor of about 2. A later modification of the original model, consisting of a two-segment lattice with a Frenkel-Kontorova (FK) on-site potential with different parameters and connected with a harmonic spring, achieved an increase in the gain to a factor of about 100--200~\cite{Li04a}, and when one of the segments was substituted with a Fermi-Pasta-Ulam (FPU) lattice, a factor of 2000 was achieved~\cite{Li05}. Following this idea of coupled systems with asymmetric properties there have been recent proposals of a solid-state device consisting of two juxtaposing materials with nonuniform thermal conductivities~\cite{Kobayashi20} and a model of an asymmetric network structure composed of two parts with different topologies~\cite{Xiong18} to obtain thermal rectification. Another theoretically proposed lattice model for controlling heat current consists of a FPU lattice with a linear mass gradient along its length~\cite{Yang07}. Although this model has a low rectification efficiency compared to the aforementioned ones, it has the advantage of being inspired by the first nanoscopic-sized experimental implementation of a thermal rectifying device constructed from a mass-loaded carbon nanotube reported to date~\cite{Chang06a}. Furthermore, since the mass-graded FPU lattice is a single-segment system, it is easier to implement and avoids altogether the problem of properly controlling the interfacial properties, which has been shown to hinder the rectification efficiency of the aforementioned two- and three-segment 1D models in the large-system-size limit~\cite{Hu06,Hu06a}. The idea of employing a graded property has been further applied to a closed billiard model with a graded magnetic field along its length~\cite{Casati07}, to a chain of elastically colliding, asymmetrically shaped mass-graded particles~\cite{Wang12}, and to a graded harmonic lattice with a quartic on-site anharmonic potential, self-consistent heat reservoirs, and weak particle interactions~\cite{Pereira10b}.

The above works demonstrate the possibility of manipulating heat flow by changing the structure and/or parameters of a given anharmonic lattice. However, these studies, among others, are focused on 1D systems. On the other hand, thermal rectification in three-dimensional (3D) systems has not been studied to the same extent as in 1D systems, although the former are closer to both experimental implementations and practical applications. Along this line a 3D extension of the aforementioned coupled FK and FPU lattice system has been performed in both in two~\cite{Lan06} and three~\cite{Lan07} dimensions, with high rectification efficiencies as in the corresponding 1D case~\cite{Li05}. Nevertheless, this system works properly as a rectifier only at low temperatures, and the results so far obtained correspond to very small system sizes. Furthermore, the parameter optimization needed to obtain high rectification values in the 3D case is not easy due to the large number of parameters involved.

In this work we perform a 3D extension of the mass-graded FPU lattice studied previously~\cite{Yang07,Romero13,Romero17a} inspired by the scalar 3D implementation of the FPU lattice wherein the validity of Fourier's law was first verified by means of nonequilibrium simulations in a 3D system without an on-site potential~\cite{Saito10} and latter verified by an equilibrium computation of the heat current autocorrelation function~\cite{Wang10}. The geometrical simplicity of the employed model allows us to explore the behavior of thermal rectification in system sizes comparable to those studied in the latter reference. Considering sufficiently large system sizes is an important factor to be taken into account because a divergent thermal conductivity was observed in previous studies of the 3D FPU lattice when both small anharmonicity values and small system sizes were considered~\cite{Shiba06,Shiba08}.

The paper is organized as follows: In Sec.~\ref{sec:Model} we present the model system as well as the employed methodology. In Sec.~\ref{sec:SA} we report our results for the rectification of the 3D FPU lattice and its dependence on various structural parameters. A discussion of the results, as well as our conclusions, is presented in Sec.~\ref{sec:Disc}.

\section{model and methodology\label{sec:Model}}

We consider a parallelepiped 3D lattice with a scalar displacement field $q_{\mathbf n}$ defined at each lattice site ${\mathbf n}=(n_1,n_2,n_3)$, where $n_1=n_2=1,\ldots,W$ and $n_3=1,\ldots,N$. The mass at a lattice site ${\mathbf n}$ of the system obeys the linear distribution $m_{\mathbf n}=\Mmax-(n_3-1)(\Mmax-\Mmin)/(N-1)$, where $\Mmax$ ($\Mmin$) is the mass of any oscillator in the leftmost (rightmost) layer of our system with the condition $\Mmax\neq\Mmin$; therefore, there is a mass gradient along the $n_3$ spatial direction of the lattice. The value $\Mmin=1$ is considered hereafter; if $\Mmax=\Mmin=1$ is taken, then the homogeneous lattice studied in Ref.~\cite{Saito10} is recovered. Two system layers, namely, those at $n_3=1$ and $N$ are connected to Langevin heat baths; therefore, the equations of motion for a given oscillator within the lattice can be written, in terms of dimensionless variables, as $\dot q_{\mathbf n}=p_{\mathbf n}/m_{\mathbf n}$ and
\ba
\dot p_{\mathbf n}&=&\sum_{\hat{\mathbf e}}\left[(q_{{\mathbf n}+\hat{\mathbf e}}-q_{\mathbf n}) + \beta (q_{{\mathbf n}+\hat{\mathbf e}}-q_{\mathbf n})^3\right] \cr
   & + & \sum_{\tilde n_3=1}^{n_{_L}}(\xi^{_L}_{\mathbf n} - \lambda_{_L}p_{\mathbf n})\delta_{{\mathbf n},\tilde{\mathbf n}} \cr
   & + & \sum_{\tilde n_3=N-n_{_R}+1}^{N}(\xi^{_R}_{\mathbf n} - \lambda_{_R}p_{\mathbf n})\delta_{{\mathbf n},\tilde{\mathbf n}},\label{EOM}
\ea
where $\tilde{\mathbf n}=(n_1,n_2,\tilde n_3)$, $\hat{\mathbf e}$ are the unitary vectors in each of the three spatial directions, $q_{\mathbf n}$ are the scalar displacements, and $p_{\mathbf n}$ are the conjugate scalar momenta. Since the contribution of the anharmonic term depends on the average temperature (see below) of the system~\cite{Li05}, it is sufficient to take a single value in all computations of the anharmonicity parameter hereafter reported, and so we chose that employed in Ref.~\cite{Saito10}, $\beta=2$. The random force $\xi^{_{L,R}}_{\mathbf n}$ at a thermostated site ${\mathbf n}$ obeys the fluctuation-dissipation relation $\langle\xi^{_{L,R}}_{\mathbf n}(t)\xi^{_{L,R}}_{\mathbf n}(t')\rangle=2\lambda_{_{L,R}}T_{_{L,R}}m_{\mathbf n}\delta(t-t')$, where $T_{_L}$ and $T_{_R}$ are the temperatures of the left and right reservoirs, respectively, with $\lambda_{_{L,R}}=1$ being the coupling strength that quantifies the interaction of the thermostated oscillator with the corresponding reservoir. Periodic boundary conditions were imposed in the first two spatial directions and fixed in the last one. The temperatures of the left and right reservoirs, in terms of the temperature difference $\Delta T=\Tl-\Tr$ and the average temperature of the system $\To=(\Tl+\Tr)/2$, were taken as $T_{_{L,R}}=T_{_0}\pm\Delta T/2$. For $\Tl>\Tr$, i.e., when the hot reservoir is connected to the heavy end of the lattice, the above equations of motion,~(\ref{EOM}), were integrated with a stochastic velocity Verlet algorithm, and after a transient time of $10^4-10^7$ dimensionless time units, the desired temporal averages were computed in the nonequilibrium steady state of the lattice for a time interval of $10^7-10^8$ with a time step of $10^{-2}$ using an in-house Fortran code running in an Intel Xeon server. The heat flux $J_{\mathbf n}$ from site ${\mathbf n}$ to site ${\mathbf n}+\hat{\mathbf e}_3$, where $\hat{\mathbf e}_3=(0,0,1)$, is given by $J_{\mathbf n}=\langle \dot q_{{\mathbf n}+\hat{\mathbf e}_3}F_{{\mathbf n},{\mathbf n}+\hat{\mathbf e}_3} \rangle$, where $F_{{\mathbf n},{\mathbf n}+\hat{\mathbf e}_3}$ is the force on oscillator ${\mathbf n}+\hat{\mathbf e}_3$ due to the oscillator at site ${\mathbf n}$ and $\langle\cdots\rangle$ the temporal average. The average current per bond for $\DT>0$ is given by
\be
J={1\over W^2 N}\sum_{n_3=n_{_L}+1}^{N-n_{_R}-1}\sum_{n_1,n_2 =1}^W J_{\mathbf n},
\ee
which in the above-mentioned configuration of the heat reservoirs is denoted $J_{+}$. The process was then repeated with $\Tl<\Tr$ ($\DT<0$), i.e., with the heat reservoirs at the two ends swapped, and likewise the heat flux computed, now denoted $J_{-}$. Then the ratio $r=J_{+}/|J_{-}|$ quantifies the rectification power of the system; $r\rightarrow1$ indicates no rectification, whereas $r\rightarrow\infty$ indicates perfect rectification.

\section{Results\label{sec:SA}}

\subsection{Rectification dependence on the system's geometry}

\begin{figure}
\includegraphics[width=0.90\linewidth,angle=0.0]{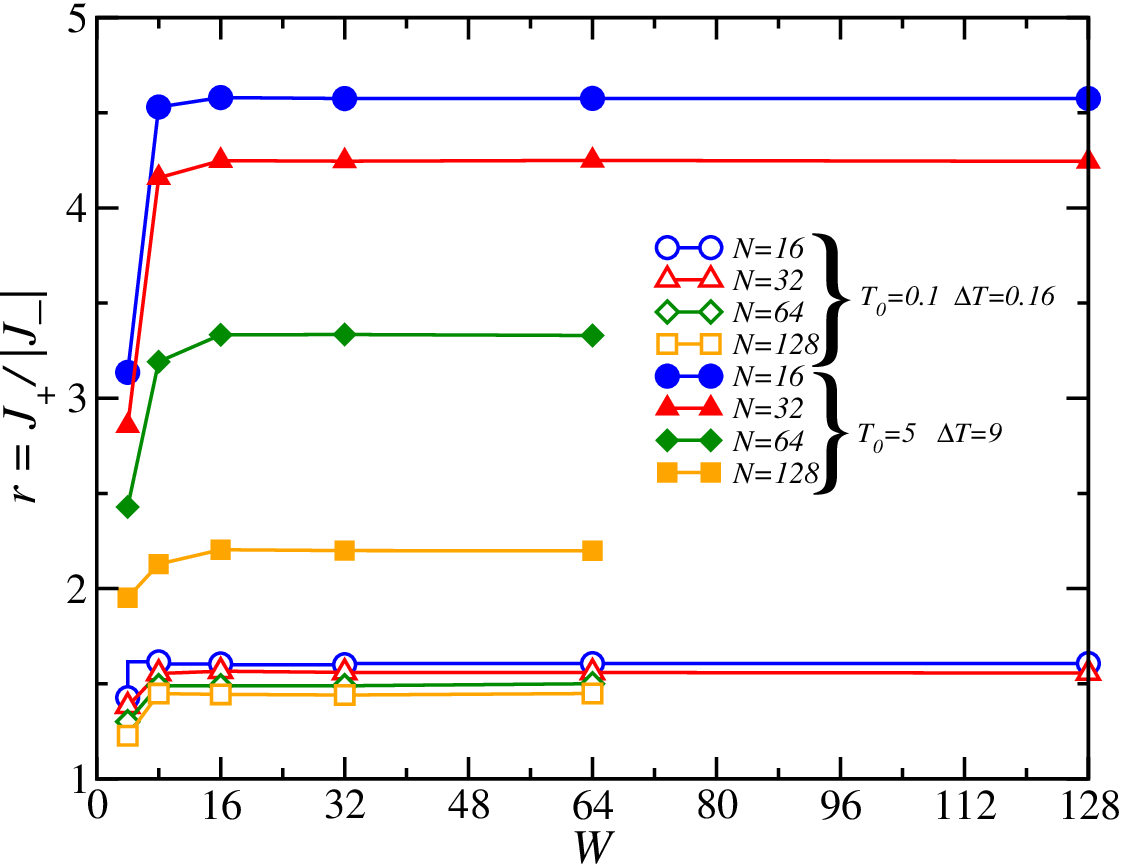}
\caption{(Color online) Thermal rectification $r$ vs system width $W$ for lateral lengths of $N=16$ (circles), $32$ (triangles), $64$ (diamonds), and $128$ (squares) with $\Mmax=10$. Open symbols correspond to $\To=0.1$ and $\DT=0.16$; filled symbols to $\To=5$ and $\DT=9$. Solid lines are a guide for the eye.}
\label{fig:uno}
\end{figure}

In Fig.~\ref{fig:uno} we plot the rectification efficiency $r$ vs the system width $W$ for various lateral system lengths $N$ and two values of the average temperature $\To$. For both temperatures it is clear that, for any fixed lateral length $N$, the value of $r$ increases as we increase $W$, but saturates quickly after a crossover width value $W_c\sim16$. This behavior is fully consistent with the already known numerical results of the heat flux $J$ in a homogeneous 3D lattice ---which decreases as $W$ increases, but then rapidly saturates to the 3D value for a very similar $W_c$--- for a homogeneous 3D FPU lattice with $\To=1.5$~\cite{Saito10}. Now, since we have explored a wider temperature range ---unlike the aforementioned reference, which employed a single average temperature value--- from our results we can infer that, for our mass-graded lattice, the saturation value of $r$ is strongly dependent on the average temperature of the system. For $\To=0.1$ rectification is insignificant, i.e., $r\sim1.5$, whereas for $\To=5$ rectification saturates at higher values, which nevertheless rapidly decrease as $N$ increases.

\begin{figure}
\includegraphics[width=0.90\linewidth,angle=0.0]{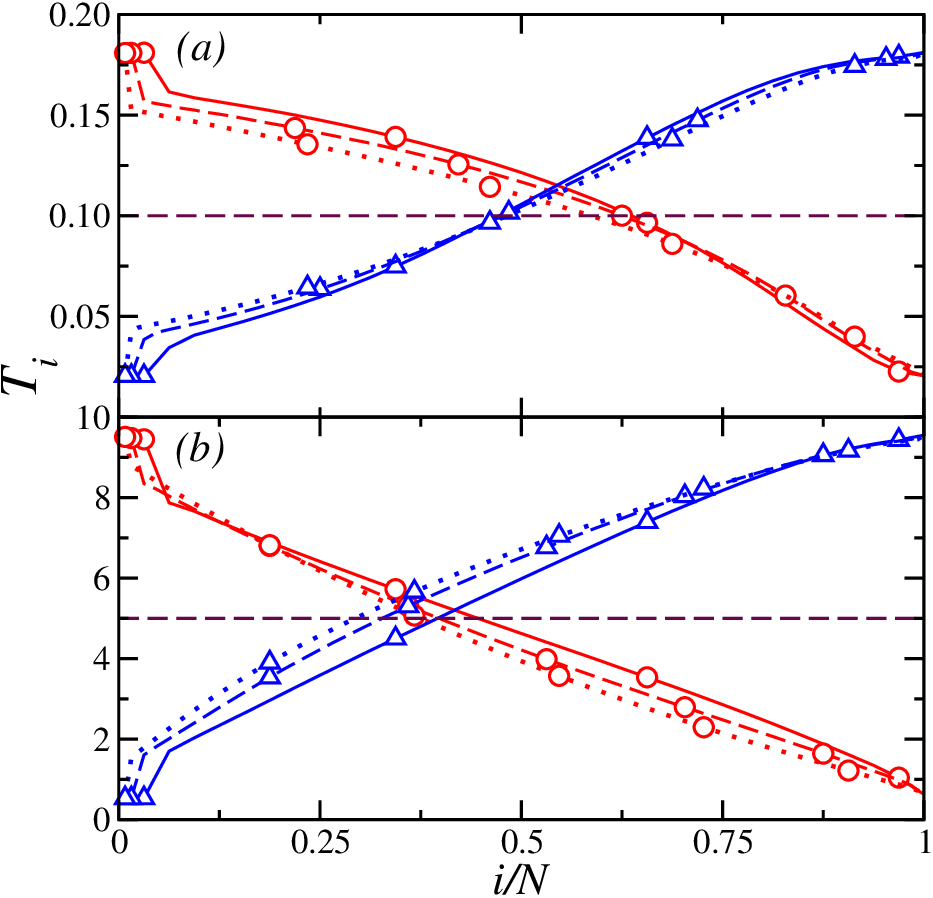}
\caption{(Color online) (a) Temperature profiles of a 3D mass-graded lattice with $W=16$, $\To=0.1$, and $\DT=0.16$; the $\Mmax$ value is the same as in Fig.~\ref{fig:uno}. Results for $\Tl>\Tr$ (circles) and $\Tl<\Tr$ (triangles) are shown. Solid, dashed, and dotted lines correspond to $N=32$, $64$, and $128$ respectively. (b) Same as (a), but for $\To=5$ and $\DT=9$. In both instances the dashed horizontal lines represent the corresponding $\To$ values.}
\label{fig:dos}
\end{figure}

The low $r$ value obtained in the $\To=0.1$ case can be inferred from the behavior of the temperature profiles for the forward- and reverse-bias configurations reported in Fig.~\ref{fig:dos} for $W=16$ and some representative $N$ values employed in Fig.~\ref{fig:uno}. For the $\To=0.1$ case reported in Fig.~\ref{fig:dos}(a) it is clear that, even though the spatial symmetry has been broken by the imposed mass gradient (especially at the left end of the system), the reflection symmetry with respect to $\To$ still holds to a large extent and largely persists for all considered $N$ values; this behavior is incompatible with a significant rectification figure. The corresponding profiles for $\To=5$ are presented in Fig.~\ref{fig:dos}(b). It is now clear that both the spatial and the reflection (around $\To$) symmetries have been to a large extent reduced, leading to the higher rectification figures reported in Fig.~\ref{fig:uno}.

\begin{figure}
\includegraphics[width=0.90\linewidth,angle=0.0]{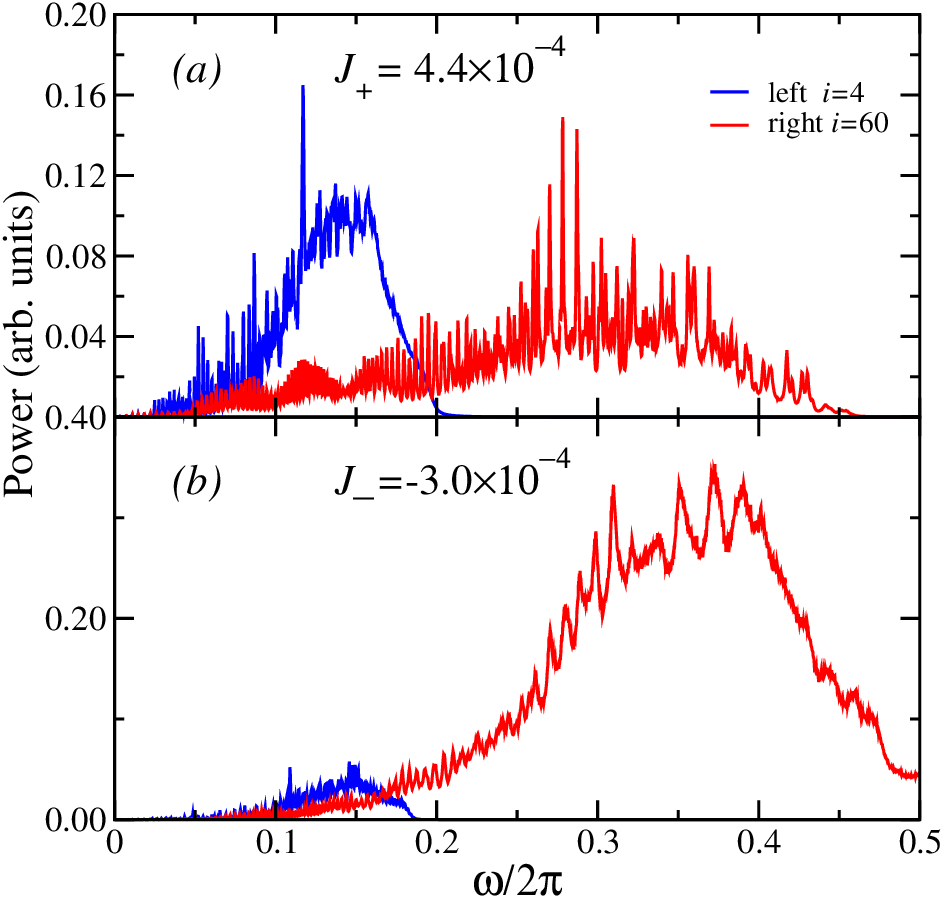}
\caption{(Color online) (a) Power spectra for an oscillator on the left side, $i=4$ (red), and one on the right side, $i=60$ (blue), for $W=16$ with $N=64$, $\Mmax=10$, $\To=0.1$, $\DT=0.16$, and $\Tl>\Tr$. (b) Same as (a), but for $\Tl<\Tr$.}
\label{fig:tres}
\end{figure}

Next we explore the relative contribution of low- and high-frequency phonons to the rectification effect under the same conditions as in Fig.~\ref{fig:dos}(a). In Fig.~\ref{fig:tres} we plot the phonon spectra $|\tau^{-1}\!\!\int_{_0}^{\tau}\!\! dt\dot q_i(t)\exp(-\mathrm{i}\omega t)|^2$ of two oscillators close to the left (heavy) and right (light) ends of the system in the low-temperature instance with $\To=0.1$. For the $J_{+}$ configuration depicted in Fig.~\ref{fig:tres}(a) the spectrum of the oscillator on the left side lies within the low-frequency region. On the other hand, the right-side spectrum is predominantly concentrated in the high-frequency region but, nevertheless, has an active low-frequency band that increases the frequency range wherein both spectra overlap, thus favoring the heat flux in the left-right direction. In both instances the distinctly discrete structure characteristic of the FPU lattice at a low temperature/energy is clearly visible. Next, for the $J_{-}$ configuration depicted in Fig.~\ref{fig:tres}(b) the contribution of the left-side spectrum is greatly diminished, whereas that of the right-side spectrum is characterized by a strong activation of high-frequency phonons. Therefore, the overlap with the left spectrum is almost suppressed, which in turn leads to the moderate decrease of $\sim30\%$ in $J_{-}$ and the low rectification figure reported in Fig.~\ref{fig:uno}.

\begin{figure}
\includegraphics[width=0.90\linewidth,angle=0.0]{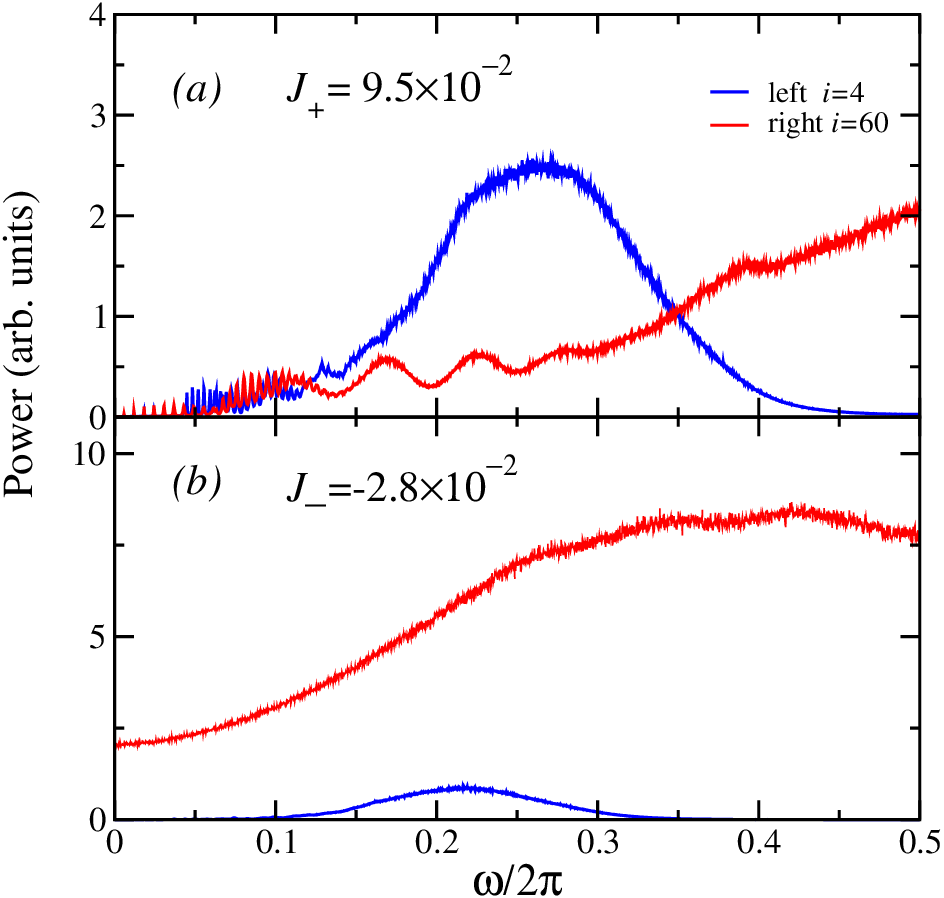}
\caption{(Color online) Same as described in the caption to Fig.~\ref{fig:tres}, but for $\To=5$ and $\DT=9$. (a) $\Tl>\Tr$ and (b) $\Tl<\Tr$.}
\label{fig:cuatro}
\end{figure}

The corresponding phonon spectra for the high-temperature regime defined by $\To=5$ are presented in Fig.~\ref{fig:cuatro}. As can be readily appreciated there are significant differences worth noting. For the forward-bias instance plotted in Fig.~\ref{fig:cuatro}(a) the left-side spectrum is shifted to higher frequencies. The contribution of the high-frequency phonons to the right-side spectrum is greatly increased with respect to the corresponding low-temperature case reported in Fig.~\ref{fig:tres}(a); furthermore, the power spectrum increases as the frequency value does so in the whole value range. The overlap of these spectra is greater than of those presented in Fig.~\ref{fig:tres}, and therefore there is an appreciable heat flow along the system. In the reverse-bias configuration [Fig.~\ref{fig:cuatro}(b)] the right-side spectrum is more concentrated around high frequencies and the contribution of each frequency in the considered value range is greater compared to the corresponding spectrum for $\To=0.1$; the left spectrum has a much smaller contribution than the corresponding one in the forward-bias configuration. These features render the overlap of both spectra insignificant, which in turns leads to the appreciable decrease of $\sim70\%$ in $J_{-}$ and a higher rectification figure than that obtained in the low-temperature case.

\subsection{Rectification dependence on model parameters and system size}

\begin{figure}
\centerline{\includegraphics*[width=80mm]{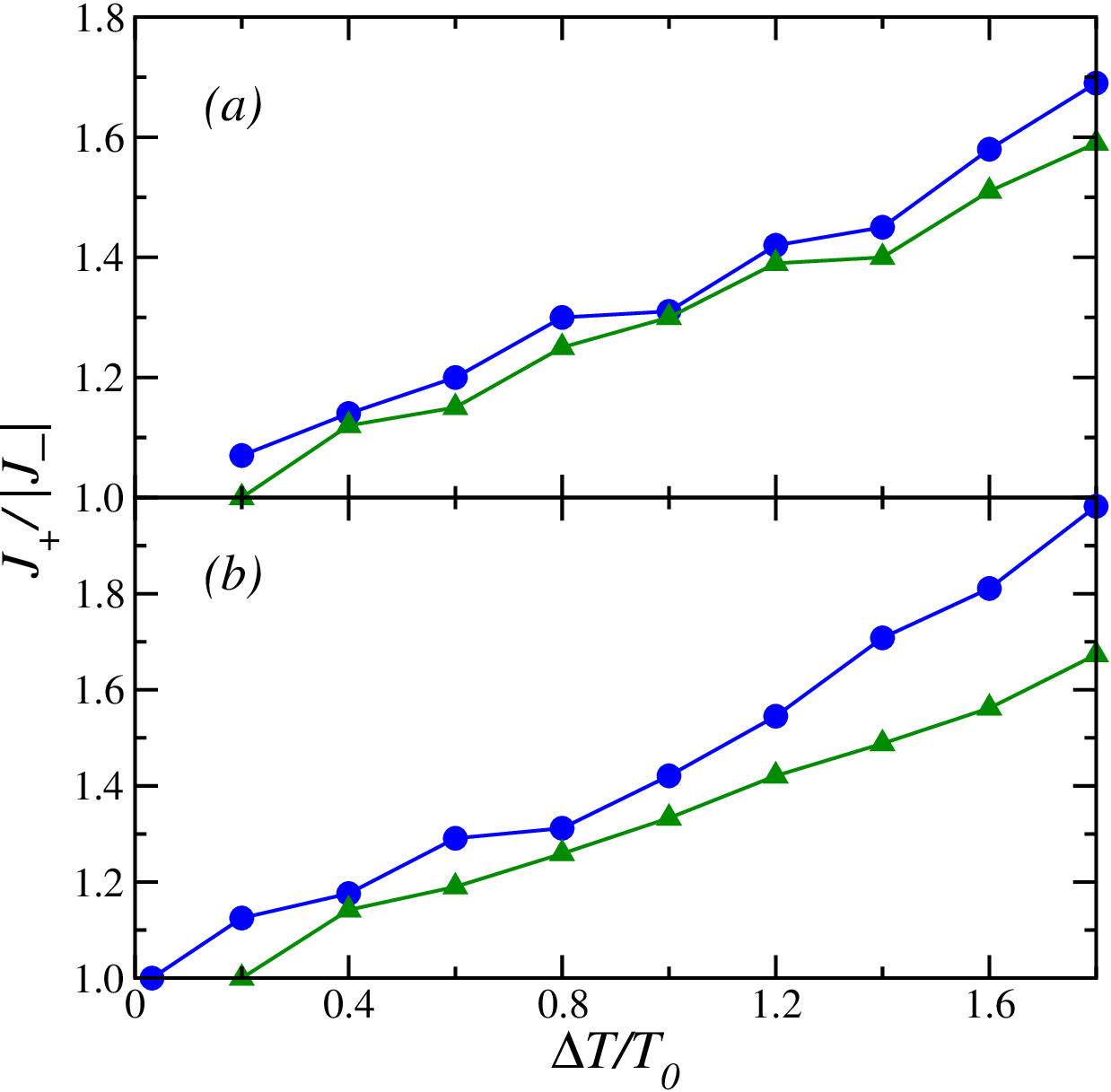}}
\caption{(Color online) (a) Dependence of thermal rectification $r$ on temperature difference $\DT$ for two lateral lengths $N$, 32 (circles) and 64 (triangles), both with $W=16$, $\Mmax=10$, and $\To=0.1$. (b) Same as (a), but for $\To=5$. Solid lines are a guide for the eye.}
\label{fig:cinco}
\end{figure}

\begin{figure}
\centerline{\includegraphics*[width=80mm]{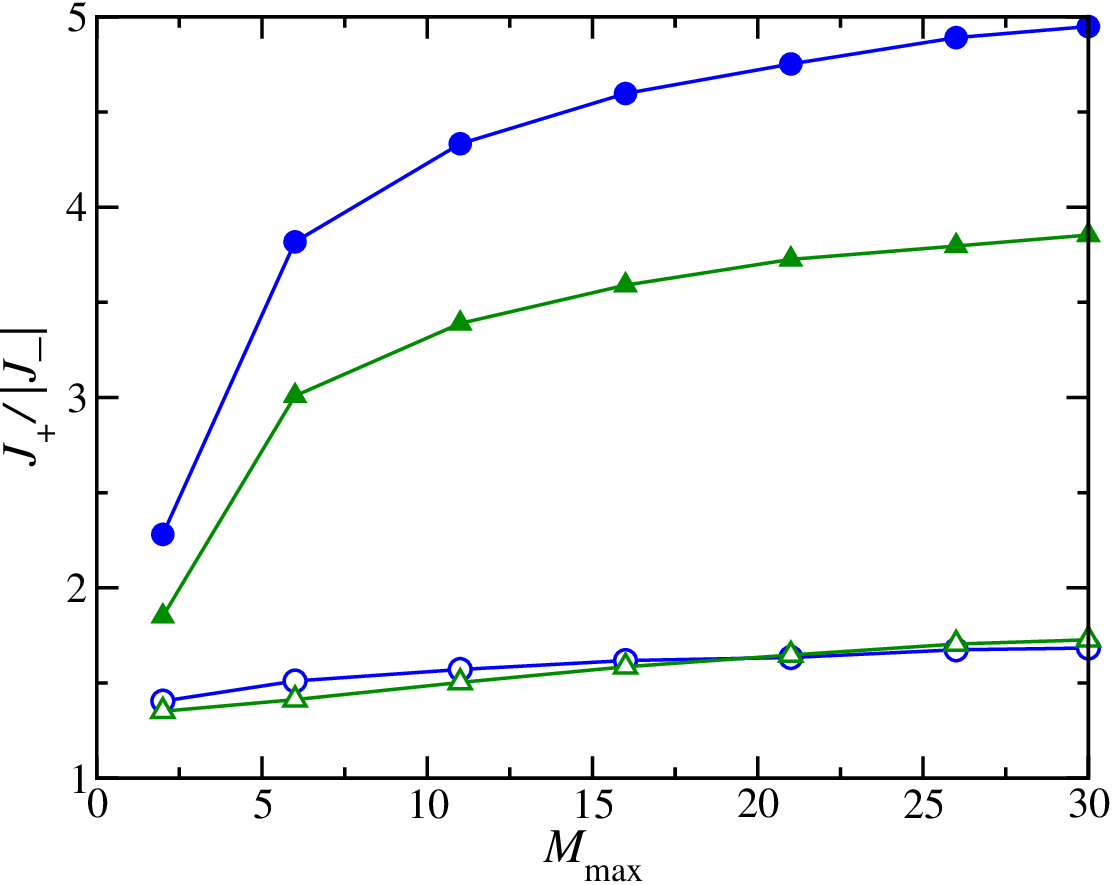}}
\caption{(Color online) (a) Thermal rectification $r$ as a function of the largest mass $\Mmax$ for $N=32$ (circles) and $N=64$ (triangles), both with $W=16$. Open symbols correspond to $\To=0.1$ and $\DT=0.16$, whereas filled ones correspond to $\To=5$ and $\DT=9$. Solid lines are a guide for the eye.}
\label{fig:seis}
\end{figure}

In the following we analyze the dependence of the rectification efficiency on other parameters that determine the behavior of the system. In Fig.~\ref{fig:cinco} we plot $r$ as a function of the temperature difference $\DT$ for the low and high average temperature values so far employed, two lateral lengths, and $\Mmax=10$. An increase in rectification is evident in both instances, independently of the considered average temperature value. In Fig.~\ref{fig:seis} we plot $r$ as a function of $\Mmax$ for low and high average temperature values. In both instances it is clear that by increasing the asymmetry of the system we obtain an increase in rectification. However, in the former case this increment is marginal, whereas in the latter it is significant; it can also be inferred that the rectification tends to saturate as the asymmetry increases, and thus, at some point, further increments in $\Mmax$ will not result in a significant increment in $r$.

\begin{figure}
\centerline{\includegraphics*[width=80mm]{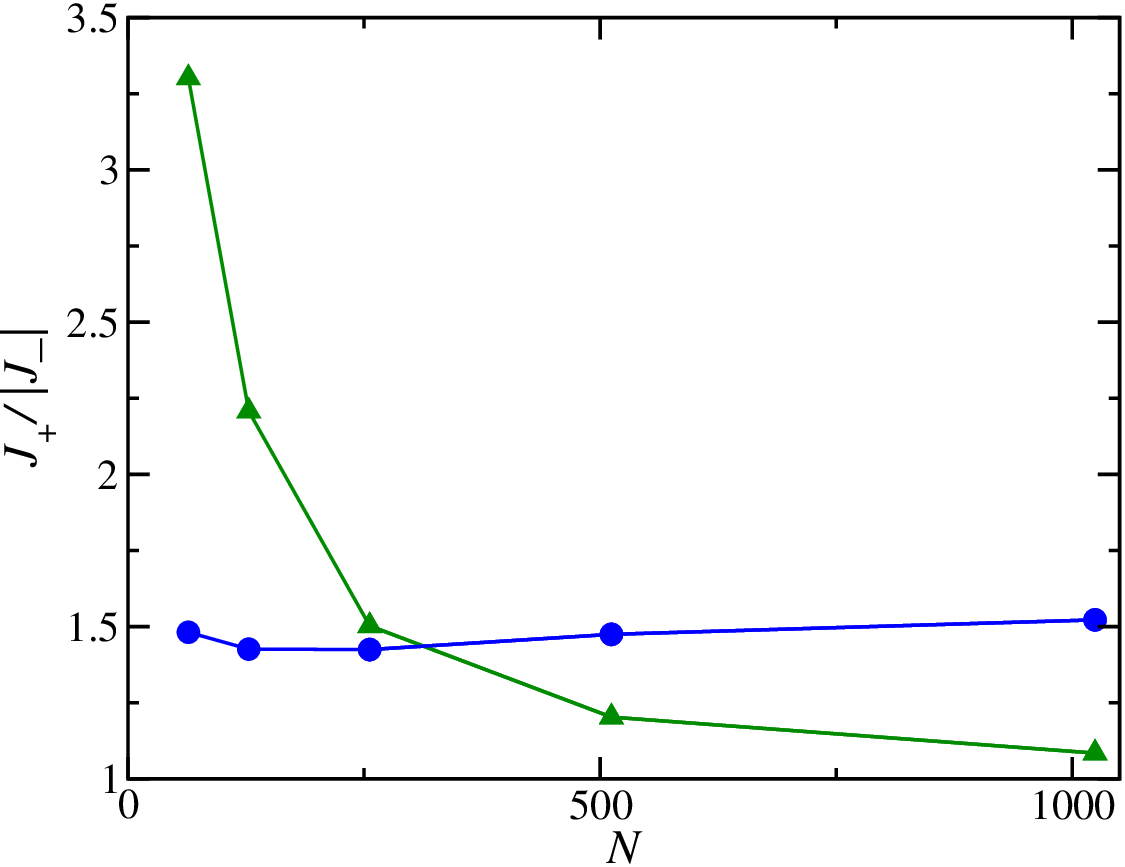}}
\caption{(Color online) Thermal rectification $r$ as a function of the lateral length $N$ for $W=16$ and $\Mmax=10$. Circles correspond to $\To=0.1$ and $\DT=0.16$; triangles, to $\To=5$ and $\DT=9$. Solid lines are a guide for the eye.}
\label{fig:siete}
\end{figure}

As for the dependence of $r$ on the lateral system size $N$, in Fig.~\ref{fig:siete} we report our results for both high and low average temperature values. In the former case the decrease in rectification is significant for $N>256$, but in the latter $r$ remains almost constant with a very small $N$ dependence over the studied value range. These results suggest that much longer simulation times would be needed to observe an appreciable decrease in rectification akin to that observed in the former case for larger $N$ values, although such large simulation times would make it very difficult to study the rectification in systems larger than those depicted in Fig.~\ref{fig:siete}. Nevertheless, with the available data it is clear that the rectification efficiency for both sets of conditions is comparable for $N<500$. 

\section{discussion and conclusions\label{sec:Disc}}

In this paper we have proposed a thermal rectifier based on a 3D mass-graded FPU lattice formerly employed to verify the validity of Fourier's law in a system without an on-site potential~\cite{Saito10}. Our first results seem to indicate that the behavior is very similar to that of the corresponding 1D lattice, especially in the fact that the obtained rectification is very small~\cite{Romero17}. It is worth noting some interesting features of the shape of the temperature profiles depicted in Fig.~\ref{fig:dos}: For $\To=0.1$ the forward-bias profile is concave downwards along the entire length of the sample, whereas the profile corresponding to the backward bias is concave downwards in the half in contact with the hot reservoir and concave upwards in the other half; this behavior indicates that the temperature gradient is nonmonotonic as a function of the position along the sample. Strong temperature jumps can be appreciated in the heavy-loaded extremes of these temperature profiles. Now, for $\To=5$ the temperature profile for the $J_{+}$ configuration is monotonic ---almost linear--- along the bulk for the smallest system size, but becomes concave upward for the larger ones. The same behavior is also observed in the $J_{-}$ configuration, except that now the temperature profile becomes concave downward as the system size is increased. Now, these results stand in sharp contrast to those previously obtained with a homogeneous 3D lattice in the forward-bias configuration, which indicate that, for $N\gg1$, the profile is indeed monotonic along the entire length of the system as in our case but becomes increasingly linear upon decreasing $\DT$~\cite{Saito10}. Our results seem to indicate that the latter features of the homogeneous lattice are rather fragile under the structural modifications and boundary conditions herein employed, which could have implications for the validity of Fourier's law for this system under the aforementioned conditions, although such verification is beyond the scope of the present work. 

It is important to recall that the best rectification values reported in Fig.~\ref{fig:uno} occur at an average temperature value $T_{_0}=5$, where no rectification is obtained in the 1D case~\cite{Yang07}. An immediate consequence is that our system, despite its low rectification power, is more robust against a deterioration of the $r$ value over a wider temperature range than its 1D counterpart, making it suitable for a larger class of possible applications~\cite{Wehmeyer17}. This result is also an advance with respect to those previously obtained with the 3D coupled FK and FPU lattice system~\cite{Lan07}, which presents no rectification whatsoever for temperatures above $T_{_0}=0.12$. Nevertheless, a readily available possibility for increasing the rectification efficiency of the 3D mass-graded lattice, especially in the low-temperature regime, could be to implement the recently proposed nonlinear system-reservoir coupling~\cite{Ming16}, which has already been shown to increase the rectification efficiency of the 1D mass-graded anharmonic FPU lattice~\cite{Romero20}.

The results in Fig.~\ref{fig:uno} seem to indicate that there is no reduction in rectifying efficiency at large $W$ values for the lateral system sizes $N$ so far considered. This behavior, derived from the fact that the heat flux attains its 3D behavior for very low $W/N$ values, is radically different from that of the 3D system composed of coupled FK and FPU lattices already studied, wherein high rectification values are obtained, but they slowly deteriorate as the perpendicular dimensions of the heat flux increase~\cite{Lan07}. From the results so far obtained it would be reasonable to assume that the system will present a nonvanishing rectification for $W/N\sim1$ values in the thermodynamic limit $N\gg1$; however, due to the results in Fig.~\ref{fig:siete}, it could also be inferred that such rectification would be exceedingly small. This reduction in $r$ with increasing system size is a recurring problem for these types of system. However, recently there have been proposals that address this problem, such as the incorporation of next-nearest-neighbor interactions~\cite{Romero17} as well as the inclusion of a ballistic channel placed between two segments (leads) defined by an on-site potential~\cite{Chen18,Romero21}, which could be easily incorporated in the model considered herein to assess their performance in reducing the aforementioned problem.

\smallskip
\begin{acknowledgments}
M.~R.~B. thanks Enrique de~la~Cuadra Coulon for useful comments and discussions and Consejo Nacional de Ciencia y Tecnolog\'\i a (CONACYT) Mexico for financial support.
\end{acknowledgments}


\bibliographystyle{prsty}

\end{document}